\documentclass[11pt]{article}
\usepackage[dvips]{graphicx}
\usepackage{mathrsfs}
\usepackage{amsthm} 
\usepackage{amsmath}
\usepackage{amssymb}
\usepackage{amsfonts}
\usepackage{float}
\usepackage{yhmath}
\usepackage{wrapfig}
\usepackage[dvips]{color}
\usepackage[dvips]{colortbl}
\usepackage{cite}
\usepackage{array}
\usepackage{tabularx}
\usepackage[sectionbib]{chapterbib}
\usepackage{threeparttable}
\usepackage{chemarrow}
\usepackage{framed}
\definecolor{shadecolor}{gray}{0.90}
\usepackage[small, bf, labelsep=quad]{caption}
\usepackage{cases}

\newcommand{\I}{I}
\newcommand{\II}{I\hspace{-0.5mm}I}

\begin{document}

\begin{center}
\Huge{Branched Polymers with Excluded Volume Effects}
\end{center}
\vspace*{-3mm}
\begin{center}
\LARGE{Relationship between Polymer Dimensions and Generation Number}
\end{center}

\vspace*{10mm}
\begin{center}
\large{Kazumi Suematsu\footnote{\, The author takes full responsibility for this article.}, Haruo Ogura$^{2}$, Seiichi Inayama$^{3}$, and Toshihiko Okamoto$^{4}$} \vspace*{2mm}\\
\normalsize{\setlength{\baselineskip}{12pt} 
$^{1}$ Institute of Mathematical Science\\
Ohkadai 2-31-9, Yokkaichi, Mie 512-1216, JAPAN\\
E-Mail: suematsu@m3.cty-net.ne.jp, ksuematsu@icloud.com  Tel/Fax: +81 (0) 593 26 8052}\\[3mm]
$^{2}$ Kitasato University,\,\, $^{3}$ Keio University,\,\, $^{4}$ Tokyo University\\[15mm]
\end{center}

\hrule
\vspace{3mm}
\noindent
\textbf{\large Abstract}: We discuss the extension of the empirical equation: $\left\langle s_{N}^{2}\right\rangle_{0}\propto g\,l^{2}$, where the subscript 0 denotes the ideal value with no excluded volume and $g$ the generation number from the root to the youngest (outermost) generation. By analogy with the linear chain problem, we introduce the assumption that the scaling relation, $\left\langle s_{N}^{2}\right\rangle_{0}\propto g^{2\lambda}\,l^{2}$, exists for arbitrary polymeric architectures, where $\lambda$ is an exponent for the backbone structure. Then, making use of the relationship between $g$ and $N$ (monomer number), we can deduce the exponent, $\nu$, for polymers with various architectures. The theory of the excluded volume effects impose the severe restriction on the quantities: $\nu_{0}$, $\nu$, and $\lambda$; for instance, the inequality, $\nu_{0}\ge\frac{1}{d+1}$, must be satisfied for isolated polymers in good solvents. An intriguing question is whether or not there exists an actual molecule that violates this inequality. We take up two examples having $\nu_{0}=1/4$ for $d=2$ and $\nu_{0}=1/6$ for $d=3$, and discuss this question.

\vspace{-2mm}

\begin{flushleft}
\textbf{\textbf{Key Words}}: Comb Polymers/ Nested Architectures/ Internal Density/ Exponent $\nu$/
\normalsize{}\\[3mm]
\end{flushleft}
\hrule
\vspace{8mm}
\setlength{\baselineskip}{14pt}

\section{Introduction}
In the preceding paper\cite{Kazumi2}, we reported that all the known (linear, star, comb, triangular, and dendrimer) polymers without the excluded volume have the mean squares of the radii of gyration that equally approach the asymptotic form:
\begin{equation}
\left\langle s_{N}^{2}\right\rangle_{0}=A\, g\,l^{2} \label{EF-1}
\end{equation}
as $g\rightarrow\infty$, where $A$ is a polymer-species-dependent coefficient and also depends on the choice of the root monomer; $g$ is the generation number from the root to the youngest (outermost) generation, and $l$ the bond length. If we accept Eq. (\ref{EF-1}) as a general rule, we can deduce the exponent, $\nu_{0}$, defined by $\left\langle s_{N}^{2}\right\rangle_{0}\propto N^{2\nu_{0}}$ ($N\rightarrow\infty$), making use of the relationship between $g$ and $N$ (total mass) alone. For instance, the regular comb polymer having side chains of the length, $n$, has the relation, $N=g+(g-1)n$ so that $N\propto g$ ($g\rightarrow\infty$), which, according to Eq. (\ref{EF-1}), yields $\nu_{0}=1/2$, whereas the triangular polymer has the relation, $N=g+\tfrac{1}{2}(g-1)(g-2)$ so that $N\propto g^{2}$, which yields $\nu_{0}=1/4$. In this way, given Eq. (\ref{EF-1}), we can deduce the exponent, $\nu_{0}$, for arbitrary polymers, without entering the intricate vectorial calculations. Then, it will be natural that one wishes to generalize Eq. (\ref{EF-1}) to predict the real exponent, $\nu$, defined by $\left\langle s_{N}^{2}\right\rangle\propto N^{2\nu}$ ($N\rightarrow\infty$).

\section{Theoretical}\label{ID-theoretical}
\begin{figure}[H]
\begin{center}
\includegraphics[width=9.5cm]{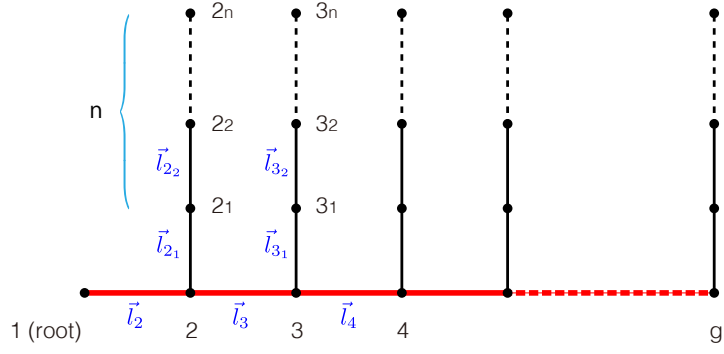} 
\caption{A comb polymer having side chains of the length, $n$.}\label{regularcomb}
\end{center}
\end{figure}
To consider the generalization of Eq. (\ref{EF-1}), let us take up the regular comb polymer\cite{Terao} (Fig. \ref{regularcomb}) having side chains of the length, $n$. This comb polymer has the total monomer number, $N=g+(g-1)n$, as mentioned above.  Let each side chain be a part of the corresponding monomer on the main backbone, and index accordingly so that, for instance, $3_{2}$ denotes the second monomers on the side chain emanating from the branching unit on the third generation, and so forth. According to Isihara\cite{Isihara, Kazumi2}, the end-to-end vector from the center of gravity to the $p$th monomer is written in the form: 
\begin{equation}
\vec{r}_{Gp}=\vec{r}_{1p}-\frac{1}{N}\sum_{p=1}^{N}\vec{r}_{1p}\label{EF-2}
\end{equation}
Following the formula (\ref{EF-2}), it is easy to show that the component vectors have the expressions:
\begin{multline}
\vec{r}_{Gh}=\frac{1}{N}\left\{\sum_{k=1}^{h-1}[N-(N-k-(k-1)n)]\,\vec{l}_{(k+1)}-\sum_{k=1}^{g-1}\left[N-k-(k-1)n\right]\,\vec{l}_{(k+1)}\right.\\
\left.-\sum_{k=1}^{g-1}\sum_{i=1}^{n}(n-i+1)\,\vec{l}_{(k+1)_{i}}+\sum_{k=1}^{h-1}[N-k-(k-1)n]\,\vec{l}_{(k+1)}\right\}\label{EF-3}
\end{multline}
\begin{multline}
\hspace{-0.5cm}\vec{r}_{Gh_{j}}=\frac{1}{N}\left\{\sum_{k=1}^{h-1}[N-(N-k-(k-1)n)]\,\vec{l}_{(k+1)}+\sum_{i=1}^{j}[N-n+i-1]\,\vec{l}_{h_{i}}-\sum_{k=1}^{g-1}\left[N-k-(k-1)n\right]\,\vec{l}_{(k+1)}\right.\\
\left.-\sum_{k=1}^{g-1}\sum_{i=1}^{n}(n-i+1)\,\vec{l}_{(k+1)_{i}}+\sum_{k=1}^{h-1}[N-k-(k-1)n]\,\vec{l}_{(k+1)}+\sum_{i=1}^{j}[n-i+1]\,\vec{l}_{h_{i}}\right\}\label{EF-4}
\end{multline}
where $1\le h\le g$ for $\vec{r}_{Gh}$; $2\le h\le g$ and $1\le j\le n$ for $\vec{r}_{Gh_{j}}$. It is seen from Eqs. (\ref{EF-3}) and (\ref{EF-4}) that the vector from the center of gravity to the $p$th monomer can generally be recast as the grand sum of all bond vectors, $\{\vec{l}_{q}\}$, that constitute the polymer,
\begin{equation}
\vec{r}_{Gp}=\frac{1}{N}\sum_{q=2}^{N} c_{q}(p)\,\vec{l}_{q}\label{EF-5}
\end{equation}
In Eq. (\ref{EF-5}), starting from $q=2$ is necessary since this polymer has $N-1$ bonds. Eq. (\ref{EF-5}) is the typical Pearson random walk\cite{Weiss, Redner} with unequal step lengths, $\{c_{q}(p)/N\}$. It is noteworthy that Eq. (\ref{EF-5}) is of the same form as the end-to-end distance of a linear chain which corresponds to the special case of $c_{q}(p)/N=1$.  Let all bonds have an equal length, $|\vec{l}_{q}|=l$, and $\vec{e}$\, be the unit vector. The mean square of the radius of gyration is written in the form:
\begin{equation}
\left\langle s_{N}^{2}\right\rangle=\frac{1}{N}\sum_{p=1}^{N}\left\langle\vec{r}_{Gp}\cdot\vec{r}_{Gp}\right\rangle=\frac{1}{N}\sum_{p=1}^{N}\left[\sum_{q=2}^{N}\frac{c_{q}(p)^{2}}{N^{2}}+\sum_{i\neq j}^{N} \frac{c_{i}(p)c_{j}(p)}{N^{2}}\left\langle\vec{e}_{i}\cdot\vec{e}_{j}\right\rangle\right]l^{2}\label{EF-6}
\end{equation}
If there is no correlation between bonds, we have $\left\langle\vec{e}_{i}\cdot\vec{e}_{j}\right\rangle=0$ for $i\neq j$. Then, Eq. (\ref{EF-6}) reduces to the freely-jointed-chain model to yield the quantity for the ideal polymer without the excluded volume. For the comb polymer under discussion, this has the form: in terms of $g$,
\begin{align}
\left\langle s_{N}^{2}\right\rangle_{0}&=\frac{1}{N}\sum_{p=1}^{N}\sum_{q=2}^{N}\left[\frac{c_{q}(p)}{N}\right]^{2}l^{2}\notag\\
&=\frac{1}{6}\cdot\frac{(n+1)^{2}g^{3}+3n^{2}(n+1)g^{2}-(n+1)(8n^{2}+2n+1)g+n(n+1)(5n+1)}{(n+1)^{2}g^{2}-2n(n+1)g+n^{2}}\,l^{2}\label{EF-7}
\end{align}
which, as $g\rightarrow\infty$, approaches the asymptotic form:
\begin{equation}
\left\langle s_{N}^{2}\right\rangle_{0}\doteq\frac{1}{6}\,g\,l^{2}\label{EF-8}
\end{equation}
the expression derived in the preceding paper\cite{Kazumi2}. If we take into account the fact that Eq. (\ref{EF-6}) has the same form as the linear chain model and that $g$ acts as the size of the molecule, the result of Eq. (\ref{EF-8}) seems to be a natural consequence. Remarkable is the fact that while, by Eqs. (\ref{EF-5}) and (\ref{EF-6}), $\left\langle s_{N}^{2}\right\rangle$ represents the statistical average of $N-1$ bonds with unequal lengths\cite{Weiss, Redner}, it behaves, phenomenologically, as though it is a linear chain linked by $g$ bonds with an equal length, $\l$.

\begin{figure}[h]
\begin{center}
\includegraphics[width=14cm]{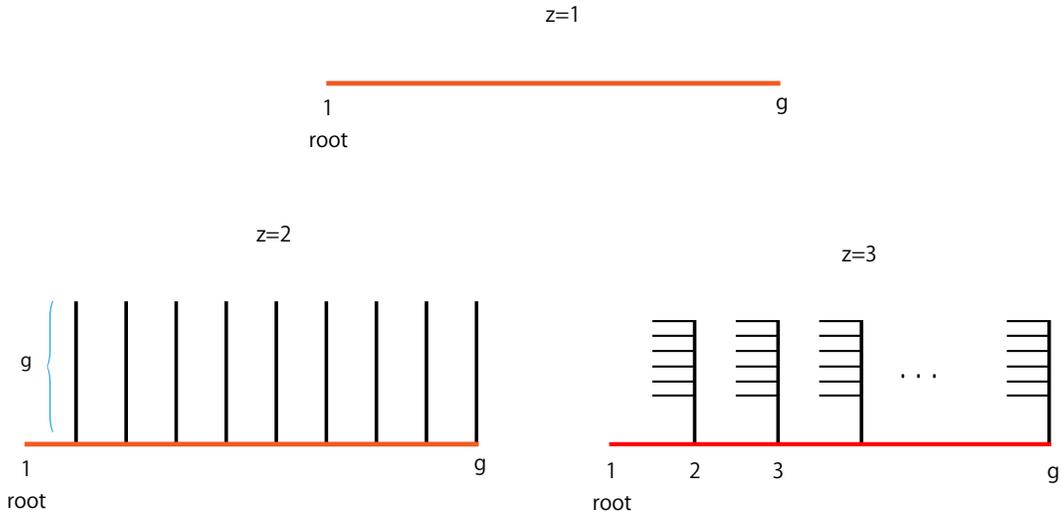} 
\caption{The first ($z=1$), the second ($z=2$), and the third ($z=3$) nested structures. In $z=3$, $g-1$ comb polymers branch off from the monomers on the new backbone (red solid-line). Such nesting may be repeated successively, which ultimately approaches the dendrimer structure with $f=3$.}\label{NestedStructure}
\end{center}
\end{figure}

For the excluded volume polymer, we have $\left\langle\vec{e}_{i}\cdot\vec{e}_{j}\right\rangle\neq 0$. In that case, by analogy with the linear chain model, it seems reasonable to expect that Eq. (\ref{EF-6}) has the asymptotic form of $g\rightarrow\infty$:
\begin{equation}
\left\langle s_{N}^{2}\right\rangle\propto g^{2\hspace{0.2mm}\lambda}\label{EF-9}
\end{equation}
where $\lambda$ is an exponent for the backbone of a molecule. To deduce the real exponent, $\nu$, let us consider the nested structures put forth in the preceding paper\cite{Kazumi2}. In Fig. \ref{NestedStructure}, the starting polymer, $N_{1}$, may be an arbitrary polymer. On each nesting, a linear polymer with $g$ monomers is newly introduced as a base polymer, on which the preceding structure is linked with each monomer on the base polymer. Such nesting is repeated successively to create deeper structures. Let $z$ represent the depth of the nest. Then, we can write generally $N_{z}=g+(g-1)N_{z-1}$. The solution to this recurrence relation is:
\begin{equation}
N_{z}=(g-1)^{z-1}N_{1}+g\frac{(g-1)^{z-1}-1}{g-2}\hspace{5mm} (z\ge 1)\label{EF-10}
\end{equation}
In this paper, we consider the simplest case in which $N_{1}$ is a linear polymer with $g$ monomers, so that $N_{1}=g$. Substituting into Eq. (\ref{EF-10}), we have $N_{z}\doteq g^{z}$ for $g\rightarrow\infty$. Substituting further this asymptotic relation into Eq. (\ref{EF-9}), we have
\begin{equation}
\left\langle s_{N}^{2}\right\rangle\propto N^{2\frac{\lambda}{z}} \hspace{10mm} (z=1, 2, 3,\dots) \label{EF-11}
\end{equation}
which yields the required relation between the exponents:
\begin{equation}
\nu=\frac{\lambda}{z} \hspace{10mm}(z=1, 2, 3,\dots)\label{EF-12}
\end{equation}
According to Eqs. (\ref{EF-1}) and (\ref{EF-9}), the ideal exponent, $\nu_{0}$, is a special case of $\lambda=1/2$ and hence
\begin{equation}
\nu_{0}=\frac{1}{2z}  \hspace{10mm} (z=1, 2, 3,\dots) \label{EF-13}
\end{equation}
From Eqs. (\ref{EF-12}) and (\ref{EF-13}), we have
\begin{equation}
\nu=2\lambda\nu_{0} \label{EF-14}
\end{equation}
Since, according to Fig. \ref{NestedStructure}, the full contour length of the backbone is, at most, $\doteq zg$, clearly $\lambda$ must satisfy $\lambda\le 1$, so that 
\begin{equation}
\nu\le 2\nu_{0} \label{EF-15}
\end{equation}
Let us apply the inequality (\ref{EF-15}) to the classic formula\cite{Flory, Issacson, Kazumi2}, the excluded volume equation for the dilution limit in good solvents:
\begin{equation}
\nu=\frac{2(1+\nu_{0})}{d+2} \label{EF-16}
\end{equation}
Substituting Eq. (\ref{EF-16}) into Eq. (\ref{EF-15}), we have
\begin{equation}
\nu_{0}\ge\frac{1}{d+1} \label{EF-17}
\end{equation}
which is a constraint condition for the classic formula (\ref{EF-16}) to have a sound physical basis\cite{Flory, Issacson, Seitz, Daoud, Kazumi2, Rosa}. While Eq. (\ref{EF-13}) indicates that there exist a large number of architectures that have different ideal exponents of the interval, $0\le\nu_{0}\le1/2$, Eq. (\ref{EF-17}) implies that, of those architectures, only those satisfying the inequality (\ref{EF-17}) are compatible with Eq. (\ref{EF-16}). For instance, $\nu_{0}\ge1/4$ is required for $d=3$\cite{Rosa}, whereas, as is seen from Eq. (\ref{EF-13}), the nesting architectures with the depth greater than $z=3$ (Fig. \ref{NestedStructure}) can not satisfy Eq. (\ref{EF-17}) since those architectures have the exponents of the interval $0\le\nu_{0}=\frac{1}{2z}\le1/6$.

\begin{figure}[h]
\begin{center}
\begin{minipage}[t]{0.46\textwidth}
\begin{center}
\includegraphics[width=6.5cm]{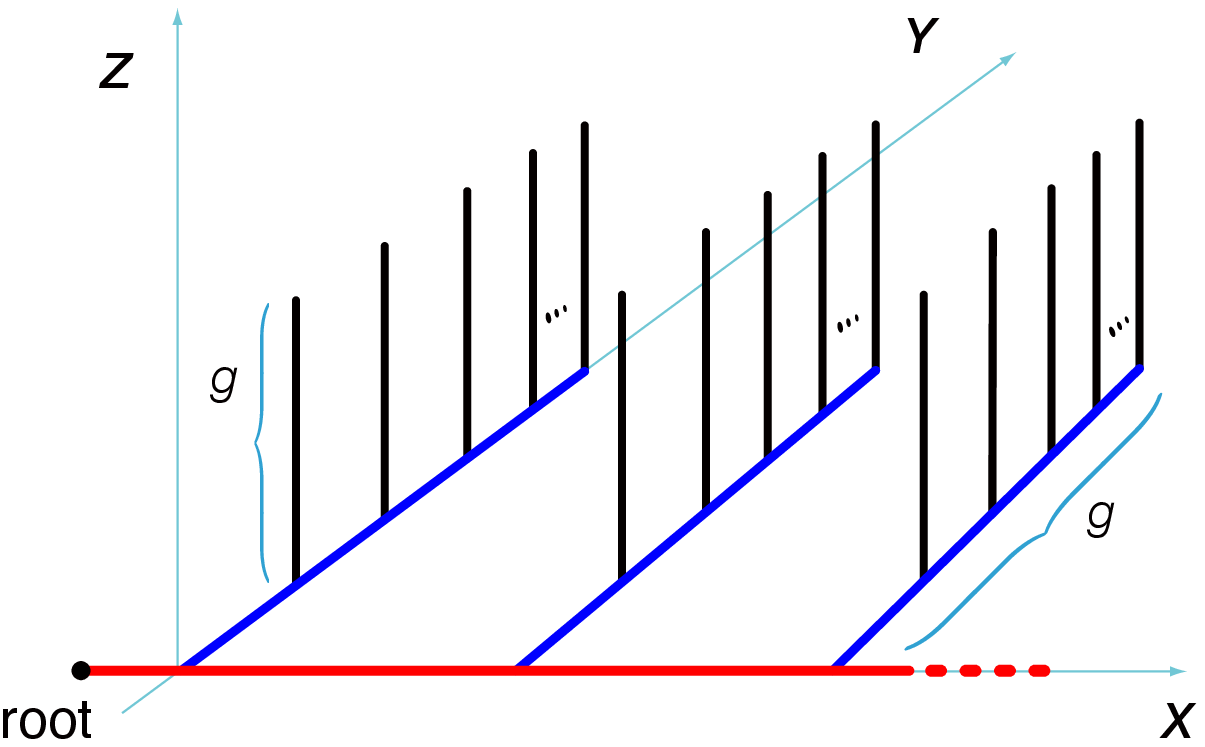}
\end{center}
\vspace{-2mm}
\caption{A simplified diagram for possible arrangement of a $z=3$ nested polymer with a large $g$ in the three-dimensional space. This polymer has $\nu_{0}=\frac{1}{6}$ and is expected to take the fully expanded conformation with $\nu=\frac{1}{3}$ in the dilution limit, according to the equations (\ref{EF-13})-(\ref{EF-15}).}\label{Comb3D}
\end{minipage}
\hspace{10mm}
\begin{minipage}[t]{0.46\textwidth}
\begin{center}
\includegraphics[width=6.5cm]{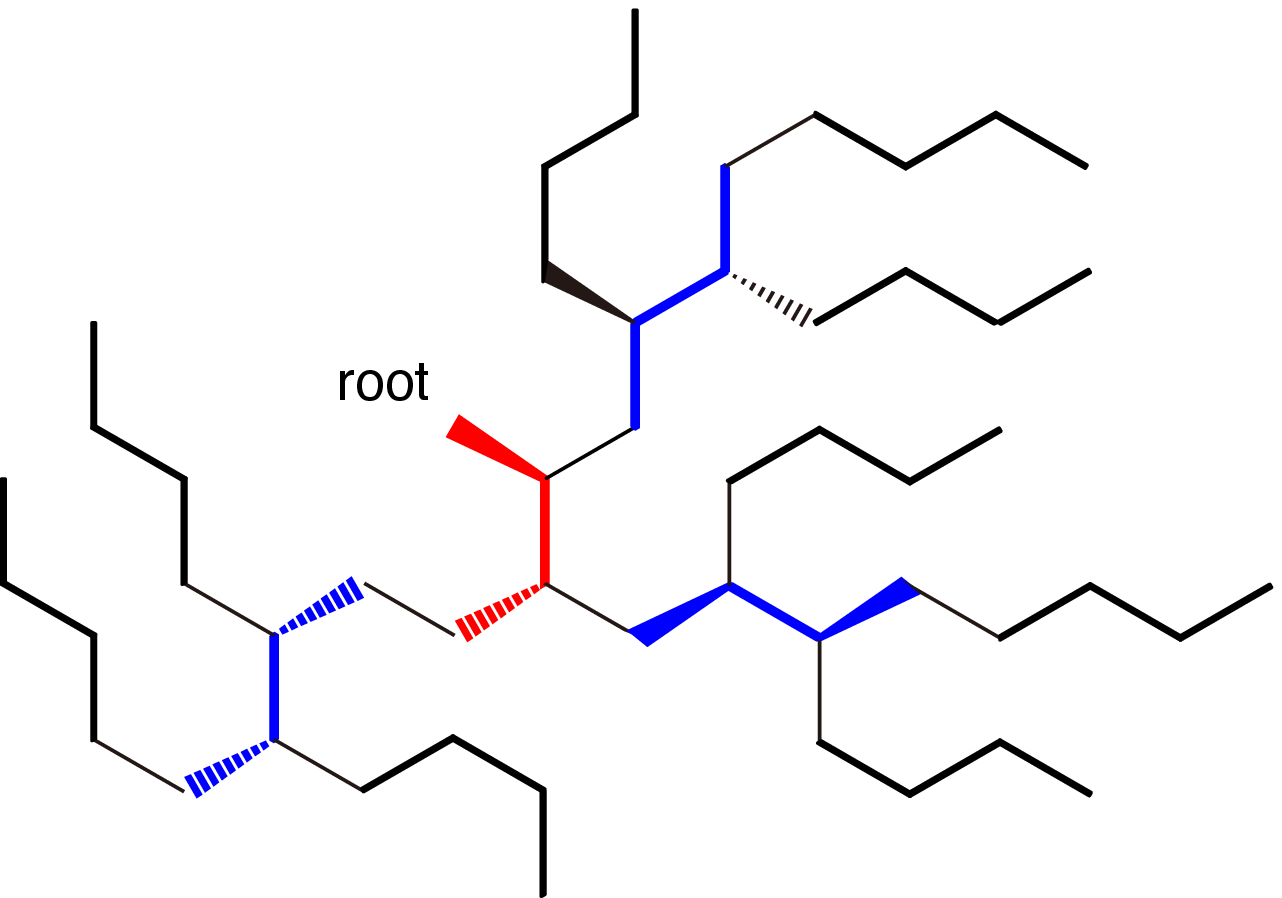}
\end{center}
\vspace{-2mm}
\caption{A $z=3$ nested polymer with $g$=4 in the three-dimensional space. The black solid-lines represent the first ($z$=1) linear molecules with $g$=4, the blue solid-lines the second ($z$=2), and the red solid-line is the third ($z$=3) backbone.}\label{z3}
\end{minipage}
\end{center}
\vspace*{-4mm}
\end{figure}

It is obvious, on the other hand, that the $z=3$ polymer with $\nu_{0}=\frac{1}{6}$ can be accommodated in the three-dimensional space, for instance, by arranging the backbone on the $XY$ plane and the combs in parallel to the $YZ$ plane so as to extend perpendicularly to each other (Fig. \ref{Comb3D}). As this example shows, in the viewpoint of synthetic chemistry, the $z=3$ polymer can be a real compound! Then, what happens to this polymer when placed in the dilution limit? One possible conjecture is that, by virtue of Eq. (\ref{EF-12}) and $\lambda=1$, $\nu$ simply takes the exponent, $1/3$, which marginally satisfies the accommodation condition, $1/d$. Under this critical circumstance, according to the thermodynamic theory of the excluded volume effects\cite{Kazumi2}, there still remains the strong Gibbs potential to enlarge the molecular dimensions due to the diffusive-flow of segments along the density gradient, which is, however, exactly canceled out by the elastic counterforce of the stretched backbone having a stronger spring constant, $\textit{\textsf{k}}$. The exponent, $1/3$, implies that the $z=3$ polymer is, in the event, rearranged into the pure liquid- or the solid-state in the limit of a large $g$. In the following section, we will examine the validity of this conjectured exponent, $\nu_{z=3}=1/3$, through the direct calculation with the help of the lattice model.

Recall that the above argument is a story for the limiting case of $N\rightarrow\infty$. A real molecule having a finite $N$ has greater freedom. Such an example of the $z$=3 polymer with $g=4$ ($N=52$) is displayed in Fig. \ref{z3}. There is a possibility that one can experimentally determine the exponent by extrapolating to $N\rightarrow\infty$, making use of samples with finite $N$'s.

\vspace*{-4mm}
\begin{center}
\begin{threeparttable}
\caption{Exponents, $\nu$, for the nesting structures in the dilution limit in good solvents ($d=3$).}\index{Square lattice}\label{R-EXP}
\begin{tabular}{l c c c c c}\hline\\[-3mm]
\hspace{5mm}Polymers & & \hspace{5mm}Exponents  & & \hspace{-0mm} \\[1.5mm]
\cline{2-6} \\[-3mm]
\hspace{5mm} & \hspace{-1mm} $z$ (depth of the nest) \hspace{-5mm} &  \hspace{2mm}$\lambda$ & \hspace{5mm} $\nu_{0}$ & \hspace{5mm} $\nu$ & \hspace{-0mm} \\[1mm]
\hline\\[-2mm]
\hspace{5mm}\textsf{Nested structures} & \hspace{0mm} $z\ge 1$ & \hspace{0mm} $\lambda\le1$ & \hspace{5mm} $\frac{1}{2z}$ & \hspace{5mm} $\frac{\lambda}{z}$ & \hspace{-0mm} \\[2mm]
\hspace{10mm} $z=1$ (linear polymer) & \hspace{0mm} 1 & \hspace{0mm} $\frac{1}{2}$ & \hspace{5mm} $\frac{1}{2}$ & \hspace{5mm} & \hspace{-0mm} \\[2mm]
\hspace{5mm} & \hspace{0mm} 1 & \hspace{0mm} $\frac{3}{5}$ & \hspace{5mm} & \hspace{5mm} $\frac{3}{5}$ & \hspace{-0mm} \\[3mm]
\hspace{10mm} $z=2$ (extended comb)\,\tnote{a} & \hspace{0mm} 2 & \hspace{0mm} $\frac{1}{2}$ & \hspace{5mm} $\frac{1}{4}$ & \hspace{5mm} & \hspace{-0mm} \\[2mm]
\hspace{10mm} & \hspace{0mm} 2 & \hspace{0mm} $1$ & \hspace{5mm} & \hspace{5mm} $\frac{1}{2}$ & \hspace{-0mm} \\[1mm]
\hspace{10mm}\cellcolor[gray]{0.9} \raisebox{-2mm}{$z=3$\,\tnote{b}} & \hspace{0mm}\cellcolor[gray]{0.9} \raisebox{-2mm}{3} & \hspace{0mm}\cellcolor[gray]{0.9} \raisebox{-2mm}{$\frac{1}{2}$} & \hspace{5mm}\cellcolor[gray]{0.9} \raisebox{-2mm}{$\frac{1}{6}$} & \hspace{5mm}\cellcolor[gray]{0.9} & \hspace{-0mm}\cellcolor[gray]{0.9} \\[4mm]
\hspace{10mm}\cellcolor[gray]{0.9} & \hspace{0mm}\cellcolor[gray]{0.9} 3 & \hspace{0mm}\cellcolor[gray]{0.9} 1 & \hspace{5mm}\cellcolor[gray]{0.9} & \hspace{5mm}\cellcolor[gray]{0.9} $\frac{1}{3}$\,\tnote{c} & \hspace{-0mm}\cellcolor[gray]{0.9} \\[2mm]
\hline\\[-7mm]
\end{tabular}
   \vspace*{2mm}
   \begin{tablenotes}
     \item a. equivalent to the comb polymer with $n=g$ in Fig. \ref{regularcomb}\cite{Kazumi2}.
     \item b. incompatible with Eq. (\ref{EF-17}), but the polymer can be embedded in the real space.
     \item c. a conjectured value according to the equations (\ref{EF-13})-(\ref{EF-15}).
   \end{tablenotes}
  \end{threeparttable}\vspace{5mm}
\end{center}

\subsection*{Two-dimensional case}
The same situation arises in the two-dimensional case of the $z=2$ (extended comb) polymer (Fig. \ref{NestedStructure}). Since a real molecule is build up based on the tetrahedral structure, a genuine two-dimensional polymer is generally unavailable. It may be useful, however, to make use of a lattice cluster\cite{Seitz, Kazumi1}, an analog of the real polymer\footnotemark. The $z=2$ cluster has $\nu_{0}=1/4$, the same value as the triangular polymer without the excluded volume, so it does not satisfy Eq. (\ref{EF-17}) in $d=2$. In contrast, according to Fig. \ref{NestedStructure}, this cluster can clearly be arranged on the two-dimensional square lattice. Thus, by Eqs. (\ref{EF-13})-(\ref{EF-15}), we expect $\lambda=1$ and $\nu=1/2$, which is again equal to the critical packing density. However, we note that the exponent, $\nu=1/2$, thus estimated is not consistent with other researchers' results that scatter from 0.5 to 0.687\cite{Issacson, Seitz, Daoud} and conflict with each other. There is a possibility that one can settle this puzzling problem through simulation experiments on lattices by using pure polymers with fixed architectures, such as the comb polymer with $n=g$ and the triangular polymer. 

In the following, we consider this problem through the direct calculation of lattice clusters, assuming that molecules are constructed from the stretched backbone and stretched side chains.

\section{Mean Squares of Radii of Gyration of Self-avoiding Polymers}\label{MRGEP}
The mean square of the radius of gyration of an excluded volume polymer can generally be evaluated by means of the thermodynamic theory\cite{Flory, Kazumi2}. Here, we evaluate the corresponding quantity in an algebraic manner, albeit for special polymers. For this purpose, we make use of the formulae for the regular comb polymer, which has the end-to-end vectors given by Eqs. (\ref{EF-3}) and (\ref{EF-4}).

We are interested in the excluded volume phenomena of polymers that violate the inequality (\ref{EF-17}). The $z$=$2$ (extended comb) polymer in $d=2$ and the $z$=$3$ polymer in $d=3$ belong to this category, as discussed above. We are going to approach this problem according to the lattice model. Suppose these polymers take the configurations with the stretched backbone and stretched side-chains extending perpendicularly to each other; in that case, it seems evident that, as $g\rightarrow\infty$, these, when arranged on the square and the cubic lattices, respectively, will cover all the lattice sites. Our aim is to calculate the radii of the gyration of such fully expanded polymers and estimate the exponent, $\nu$. The main point of the calculation is that only the scalar products between the bond vectors parallel to each other survive, and all the perpendicular components vanish.

\subsection{The $z=1$ (Rod) Polymer in the One-Dimensional Space}
The one-dimensional freely-jointed polymer has $\nu_{0}=1/2$, and hence, it marginally satisfies the inequality (\ref{EF-17}). The end-to-end distance vector, $\vec{r}_{Gh}$, of the one-dimensional polymer corresponds to the case of $n=0$ in Eq. (\ref{EF-3}), so that only the scalar product, $\vec{l}_{k}\cdot\vec{l}_{h}$, between bonds on the backbone is mathematically useful. Let this polymer be comprised of $g$ monomers, and all the bonds have the equal length, $|\vec{l}_{k}|=l$. According to the definition of the mean square of the radius of gyration and Eq. (\ref{EF-3}), we have
\begin{equation}
\left\langle s_{g}^{2}\right\rangle_{z=1}=\frac{1}{N}\left\langle\sum_{h=1}^{g}\vec{r}_{Gh}\cdot\vec{r}_{Gh}\right\rangle=\frac{1}{12}\left(g^{2}-1\right)l^{2} \label{EF-18}
\end{equation}
For $g\rightarrow\infty$, this gives $\left\langle s_{g}^{2}\right\rangle_{z=1}\doteq \frac{1}{12}\,g^{2}\,l^{2} $. Since $N=g$, we have $\nu=1$, in accord with what is predicted by the equations (\ref{EF-13})-(\ref{EF-15}), and also agrees with the prediction by Eq. (\ref{EF-16}).

\subsection{The $z=2$ (Extended Comb) Polymer in the Two-Dimensional Space}
In Eqs. (\ref{EF-3}) and (\ref{EF-4}), we put $n=g$. With $\vec{l}_{h}\cdot\vec{l}_{k_{j}}=0$ in mind, we have
\begin{equation}
\left\langle s_{g}^{2}\right\rangle_{z=2}=\frac{1}{N}\left\{\left\langle\sum_{h=1}^{g}\vec{r}_{Gh}\cdot\vec{r}_{Gh}\right\rangle+\left\langle\sum_{h=2}^{g}\sum_{j=1}^{g}\vec{r}_{Gh_{j}}\cdot\vec{r}_{Gh_{j}}\right\rangle\right\}=\frac{1}{6}\frac{\left(g^{2}-1\right)\left(g^{2}+3\right)}{g^{2}}l^{2} \label{EF-19}
\end{equation}
which, as $g\rightarrow\infty$, reduces to $\left\langle s_{g}^{2}\right\rangle_{z=2}\doteq \frac{1}{6}\,g^{2}\, l^{2}$. Since $N=g^{2}$ for the $z$=2 polymer, we have $\nu=1/2$, in agreement with the prediction of the equations (\ref{EF-13})-(\ref{EF-15}), and also with the critical packing density, $1/d$.
\begin{shaded}
\vspace{-4mm}
\subsection*{Mathematical Check}
For $g=1$ ($N=1$), Eq. (\ref{EF-19}) gives the obvious result: $\left\langle s_{g}^{2}\right\rangle_{z=2}=0$. For $g=2$ ($N=4$), the molecule has the configuration (\raisebox{-0.2mm}{\includegraphics[width=5mm]{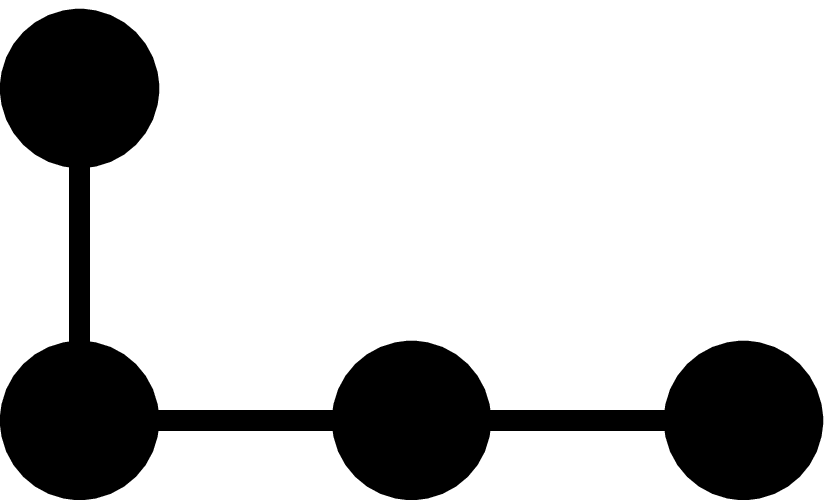}}) like the capital letter, L. It is easy to calculate, by the elementary geometry, the mean square of the radius of gyration of this molecule to yield $\left\langle s_{g}^{2}\right\rangle_{z=2}=\frac{7}{8}\,l^{2}$, in exact agreement with the prediction of Eq. (\ref{EF-19}).
\end{shaded}

\subsection{The $z=3$ Polymer in the Three-Dimensional Space}
To solve this problem, it is necessary for us to make an extension of Eqs. (\ref{EF-3}) and (\ref{EF-4}); the result is shown in Appendix \ref{Appendix A}. Using the extended equations (\ref{A2})-(\ref{A4}), we can calculate, in quite the same manner, the corresponding quantity for the $z$=3 polymer to obtain
\begin{align}
\left\langle s_{g}^{2}\right\rangle_{z=3}&=\frac{1}{N}\left\{\left\langle\sum_{h=1}^{g}\vec{r}_{Gh}\cdot\vec{r}_{Gh}\right\rangle+\left\langle\sum_{h=2}^{g}\sum_{j=1}^{g}\vec{r}_{Gh_{j}}\cdot\vec{r}_{Gh_{j}}\right\rangle+\left\langle\sum_{h=2}^{g}\sum_{j=2}^{g}\sum_{p=1}^{g}\vec{r}_{Gh_{jp}}\cdot\vec{r}_{Gh_{jp}}\right\rangle\right\}\notag\\[2mm]
&=\frac{1}{12}\frac{(g-1)\left(3g^{5}-5g^{4}+16g^{3}+6g^{2}+g-5\right)}{\left(g^{2}-g+1\right)^{2}}\,l^{2} \label{EF-20}
\end{align}
which, as $g\rightarrow\infty$, reduces to $\left\langle s_{g}^{2}\right\rangle_{z=3}\doteq \frac{1}{4}\,g^{2}\, l^{2}$. Since, by Eq. (\ref{EF-10}), $N\doteq g^{3}$ for the $z$=3 polymer, we have $\left\langle s_{N}^{2}\right\rangle_{z=3}\doteq\frac{1}{4}\,N^{\frac{2}{3}}\,l^{2}$ to yield $\nu=1/3$, in agreement with the conjectured value in Table \ref{R-EXP}.

An essential point of the above calculation results (\ref{EF-18})-(\ref{EF-20}) for the fully expanded polymers is that the mean squares of the radii of gyration are proportional to $g^{2}$, namely, $\langle s_{g}^{2}\rangle\propto g^{2}$, in contrast to the case of the ideal branched polymers that obey $\langle s_{g}^{2}\rangle_{0}\propto g$. In summary, the empirical rule might be generalized to
\begin{equation}
\langle s_{g}^{2}\rangle\propto
\begin{cases}
\hspace{1mm}g & \mbox{(for freely jointed molecules)}\\[3mm]
\hspace{1mm}g^{2} & \mbox{(for fully expanded molecules)}
\end{cases}\label{EF-21}
\end{equation}
The results are in harmony with the initial scaling assumption, $\left\langle s_{N}^{2}\right\rangle\propto g^{2\hspace{0.2mm}\lambda}$, introduced in Eq. (\ref{EF-9}). From these examples, and taking dimensional symmetry into consideration, the relations in Eq. (\ref{EF-21}) are probably a general rule and holds irrespective of the difference in architectures.

The exponent, $\nu=1/2$, deduced above for the lattice analog of the isolated $z$=2 (extended comb) polymer conflicts with the thermodynamic theory, 5/8\cite{Issacson, Daoud, Kazumi2}, the simulation experiments, $\simeq 0.61$\cite{Seitz}, on the square lattice, and the other theoretical methods, for the randomly branched polymer with the same ideal exponent $\nu_{0}=1/4$. We notice that, because of the constraint condition (\ref{EF-17}), all those approaches are not applicable to the polymers such as the $z$=2 polymer ($\nu_{0}=1/4$) in the two-dimensional space and the $z$=3 polymer ($\nu_{0}=1/6$) in the three-dimensional space. The real exponent, $\nu$, in the isolated state of these special polymers that violate the inequality (\ref{EF-17}) can be evaluated by the use of Eq. (\ref{EF-14}); namely, for polymers that obey $\nu_{0}<\frac{1}{d+1}$, we have $\lambda=1$, so it must be that
\begin{equation}
\nu=2\nu_{0}\label{EF-22}
\end{equation}
as has been fortified in Eq. (\ref{EF-19}) for the $z$=2 polymer, and in Eq. (\ref{EF-20}) for the $z$=3 polymer.

In spite of the advent of such unprecedented polymers that violate the constraint condition (\ref{EF-17}), the fundamental equation of force (\ref{EF-23}) for the excluded volume phenomena is invariably valid for almost all actual polymers synthesized in laboratories.
\begin{equation}
\frac{d G}{d\alpha}=\left(\mu_{c_{2\II}}-\mu_{c_{2I}}\right)\frac{d c_{2\II}}{d\alpha}+\frac{dG_{\text{elasticity}}}{d\alpha}=0\label{EF-23}
\end{equation}
(In Eq. (\ref{EF-23}), the subscript 2 denotes polymer segments; the subscripts $\I$ and $\II$ a more dilute region and a more concentrated region, respectively.)

\section{Discussion}
Through the previous to the present papers, we have shown that there are an infinite number of architectures having different exponents, $\nu_{0}$. We showed that, of those architectures, only those that satisfy the inequality (\ref{EF-17}) are compatible with the classic formula\cite{Issacson}. We have investigated two examples that deviate from the inequality (\ref{EF-17}): one is the $z=2$ nested structure ($\nu_{0}=1/4$) in $d=2$, and the other is the $z=3$ nested structure ($\nu_{0}=1/6$) in $d=3$. Despite the deviation from the inequality (\ref{EF-17}), for instance, the $z=2$ polymer in $d=2$ is expected to have a reality, because it can be embedded in the two-dimensional space. We showed that the same is true for the $z=3$ nested structure in $d=3$. An intriguing aspect is that the mean squares of the radii of gyration obey the relationship: $\left\langle s_{g}^{2}\right\rangle\propto g^{2}$ for the fully expanded polymers. From the viewpoint of the dimensional symmetry, there is a reason to conjecture that this relationship is  independent of architectures.

\footnotetext{\, We want to call attention to the fact that, if configurational isomerization is carried out according to an algorithm during a simulation to deal with, for instance, the randomly branched polymers, the lattice clusters necessarily give rise to different architectures and hence different isomer ratios, from the corresponding chemical molecules to be produced under the assumption of the equal reactivity of functional units. This is because of the realization of the steric hindrance in lattice clusters that leads to the reduction of the functionality, $f$ \cite{Kazumi1}. This aspect has early been pointed out by Klein\cite{Seitz}.}

An important fact clarified through this work is that the thermodynamic theory (\ref{EF-23}) of the excluded volume effects\cite{Flory, Kazumi2} and the scaling formula (\ref{EF-14}) complement each other and are, by no means, exclusive. Eq. (\ref{EF-23}) is useful for the architectures that satisfy $\nu_{0}\ge\frac{1}{d+1}$ (most of the real polymers in $d=3$ belong to this category!), while Eq. (\ref{EF-14}) is useful in predicting the exponent, $\nu$, of the architectures that fulfill $\nu_{0}\le\frac{1}{d+1}$, which necessarily leads to $\lambda=1$ and $\nu=2\nu_{0}$, under the condition of the dilution limit in good solvents. Whereas the scaling formula is a law restricted to $N\rightarrow\infty$, the thermodynamic theory covers a broader range of $N$ and a broader range of polymer concentration from the dilution limit to the melt; it is useful in extracting the physicochemical facets of the excluded volume phenomena.

\vspace{10mm}
\renewcommand{\appendixname}{Appendix}

\appendix

\section*{\LARGE \textbf{Appendix}}
\vspace*{5mm}
\numberwithin{equation}{section}
\setcounter{equation}{0}
\renewcommand{\thefigure}{A\arabic{figure}}
\setcounter{figure}{0}

\section{Spatial Configuration of the $z=3$ Polymer}\label{Appendix A}
We calculate the spatial configuration and the mean square of the radius of gyration of this polymer, which can be accomplished with the help of the Isihara formula\cite{Isihara, Kazumi2}:
\begin{figure}[h]
\vspace*{5mm}
\begin{center}
\includegraphics[width=9.5cm]{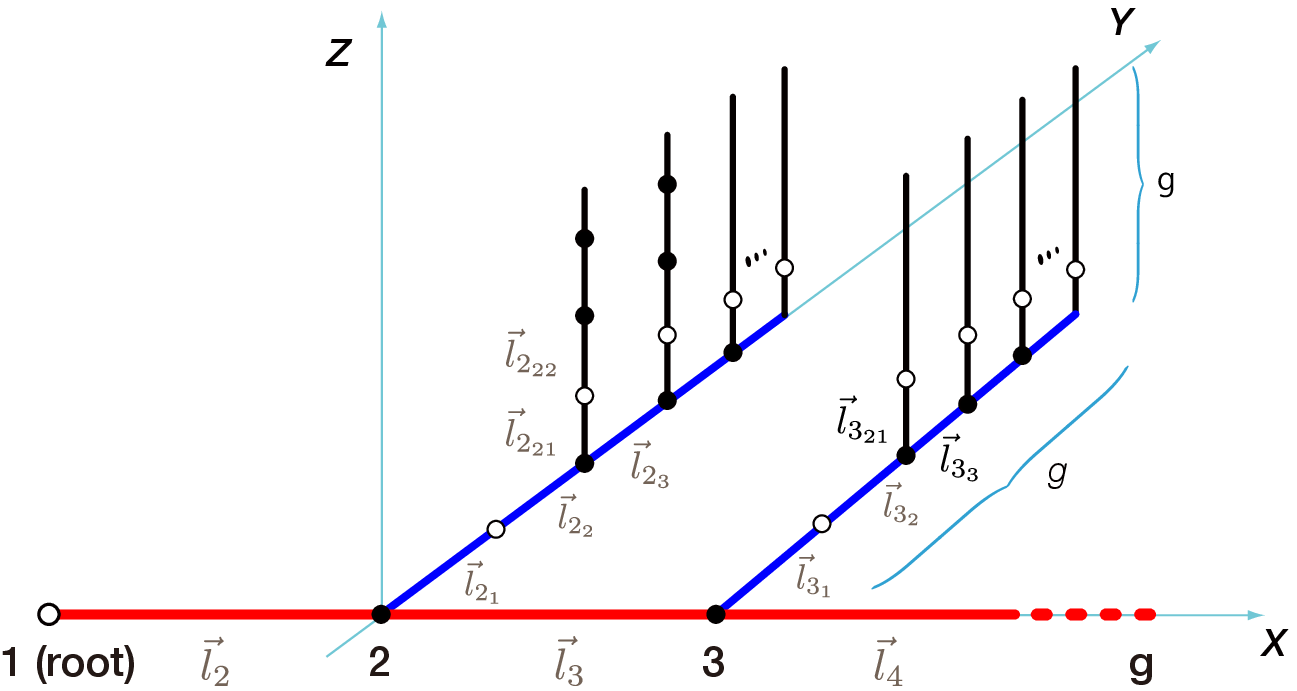} 
\caption{The $z$=3 polymer having the stretched backbone and the stretched side chains of the length, $g$.}\label{z=3polymer}
\end{center}
\end{figure}

\begin{equation}
\vec{r}_{Gp}=\vec{r}_{1p}-\frac{1}{N}\sum_{p=1}^{N}\vec{r}_{1p}\label{A1}
\end{equation} 
The key point is to rearrange the end-to-end vector, $\vec{r}_{Gp}$, from the center of mass to the $p$th monomer into the grand sum of all the bond vectors that constitute the polymer: the result being
\begin{multline}
\vec{r}_{Gh}=\frac{1}{N}\left\{\sum_{k=1}^{h-1}[N-(N-k-(k-1)g^{2})]\,\vec{l}_{(k+1)}-\sum_{k=1}^{g-1}\left[N-k-(k-1)g^{2}\right]\,\vec{l}_{(k+1)}\right.\\
-\sum_{k=1}^{g-1}\left(g^{2}\,\vec{l}_{(k+1)_{1}}+\sum_{i=2}^{g}\left\{g^{2}-[i-1+(i-2)g]\right\}\,\vec{l}_{(k+1)_{i}}\right)-\sum_{k=1}^{g-1}\sum_{i=2}^{g}\sum_{p=1}^{g}(g-p+1)\vec{l}_{(k+1)_{ip}}\\
\left.+\sum_{k=1}^{h-1}\left[N-k-(k-1)g^{2}\right]\,\vec{l}_{(k+1)}\right\}\label{A2}
\end{multline}
\begin{multline}
\vec{r}_{Gh_{j}}=\frac{1}{N}\Bigg\{\sum_{k=1}^{h-1}[N-(N-k-(k-1)g^{2})]\,\vec{l}_{(k+1)}+\left(N-g^{2}\right)\vec{l}_{h_{1}}+\sum_{i=2}^{j}\left[N-\left(g^{2}-[i-1+(i-2)g]\right)\right]\,\vec{l}_{h_{i}}\\
-\sum_{k=1}^{g-1}\left[N-k-(k-1)g^{2}\right]\,\vec{l}_{(k+1)}-\sum_{k=1}^{g-1}\left(g^{2}\,\vec{l}_{(k+1)_{1}}+\sum_{i=2}^{g}\left\{g^{2}-[i-1+(i-2)g]\right\}\,\vec{l}_{(k+1)_{i}}\right)\\
-\sum_{k=1}^{g-1}\sum_{i=2}^{g}\sum_{p=1}^{g}(g-p+1)\vec{l}_{(k+1)_{ip}}+\sum_{k=1}^{h-1}\left[N-k-(k-1)g^{2}\right]\,\vec{l}_{(k+1)}\\
+g^{2}\,\vec{l}_{h_{1}}+\sum_{i=2}^{j}\left(g^{2}-[i-1+(i-2)g]\right)\,\vec{l}_{h_{i}}\Bigg\}\label{A3}
\end{multline}
\begin{multline}
\vec{r}_{Gh_{jp}}=\frac{1}{N}\Bigg\{\sum_{k=1}^{h-1}[N-(N-k-(k-1)g^{2})]\,\vec{l}_{(k+1)}+\left(N-g^{2}\right)\vec{l}_{h_{1}}+\sum_{i=2}^{j}\left[N-\left(g^{2}-[i-1+(i-2)g]\right)\right]\,\vec{l}_{h_{i}}\\
+\sum_{q=1}^{p}[N-(g-q+1)]\,\vec{l}_{h_{jq}}-\sum_{k=1}^{g-1}\left[N-k-(k-1)g^{2}\right]\,\vec{l}_{(k+1)}\\
-\sum_{k=1}^{g-1}\left(g^{2}\,\vec{l}_{(k+1)_{1}}+\sum_{i=2}^{g}\left\{g^{2}-[i-1+(i-2)g]\right\}\,\vec{l}_{(k+1)_{i}}\right)-\sum_{k=1}^{g-1}\sum_{i=2}^{g}\sum_{q=1}^{g}(g-q+1)\vec{l}_{(k+1)_{iq}}\\
+\sum_{k=1}^{h-1}\left[N-k-(k-1)g^{2}\right]\,\vec{l}_{(k+1)}+g^{2}\,\vec{l}_{h_{1}}+\sum_{i=2}^{j}\left(g^{2}-[i-1+(i-2)g]\right)\,\vec{l}_{h_{i}}+\sum_{q=1}^{p}(g-q+1)\,\vec{l}_{h_{jq}}\Bigg\}\label{A4}
\end{multline}
where $\vec{l}_{h_{jp}}$ represents the $p$th bond vector emanating from $j$th monomer on the side chain which emanates from $h$th monomer on the backbone (the red solid line in Fig. \ref{z=3polymer}). In Eqs. (\ref{A2})-(\ref{A4})
\begin{eqnarray}
\bullet & 1\le h\le g & \hspace{3mm}\text{for}\hspace{4mm} \vec{r}_{Gh}\notag\\
\bullet & 2\le h\le g \hspace{3mm} \text{and}\hspace{3mm}  1\le j\le g &\hspace{3mm}\text{for}\hspace{4mm} \vec{r}_{Gh_{j}}\label{A5}\\
\bullet & 2\le h\le g, \hspace{3mm} 2\le j\le g, \hspace{3mm} \text{and}\hspace{3mm} 1\le p\le g &\hspace{3mm}\text{for}\hspace{4mm}\vec{r}_{Gh_{jp}}\notag
\end{eqnarray}

First, let us consider the fully expanded molecule. Assume that, in the limit of a large $g$, this polymer takes the configuration with the stretched backbone and the stretched side chains (Fig. \ref{z=3polymer}). Then, the polymer can marginally be arranged on the simple cubic lattice with the side length $l$. Since the backbone and the side chains extend perpendicularly to each other, all the scalar products between the perpendicular bond-vectors should vanish, and hence
\begin{align}
\left\langle s_{g}^{2}\right\rangle_{z=3}&=\frac{1}{N}\left\{\left\langle\sum_{h=1}^{g}\vec{r}_{Gh}\cdot\vec{r}_{Gh}\right\rangle+\left\langle\sum_{h=2}^{g}\sum_{j=1}^{g}\vec{r}_{Gh_{j}}\cdot\vec{r}_{Gh_{j}}\right\rangle+\left\langle\sum_{h=2}^{g}\sum_{j=2}^{g}\sum_{p=1}^{g}\vec{r}_{Gh_{jp}}\cdot\vec{r}_{Gh_{jp}}\right\rangle\right\}\notag\\[2mm]
&=\frac{1}{12}\frac{(g-1)\left(3g^{5}-5g^{4}+16g^{3}+6g^{2}+g-5\right)}{\left(g^{2}-g+1\right)^{2}}\,l^{2} \label{A6}
\end{align}
As $g\rightarrow\infty$, Eq. (\ref{A6}) reduces to
\begin{equation}
\left\langle s_{g}^{2}\right\rangle_{z=3}\doteq\frac{1}{4}\,g^{2}\,l^{2}\label{A7}
\end{equation}
Applying the relation (\ref{EF-10}), $N=g(g^{2}-g+1)$, in the text, we have $\left\langle s_{N}^{2}\right\rangle_{z=3}\doteq\frac{1}{4}\,N^{\frac{2}{3}}\,l^{2}$. Thus,
\begin{equation}
\nu_{z=3}=\frac{1}{3}\hspace{5mm}(d=3)\label{A8}
\end{equation}
for the fully expanded molecule. The result is in accord with the conjectured value in the text (Table \ref{R-EXP}).

\begin{shaded}
\vspace{-3mm}
\subsection*{Mathematical Check}
Apply $g=2$ ($N=6$) to Eq. (\ref{A6}), and we have: $\left\langle s_{N}^{2}\right\rangle_{z=3}=\frac{55}{36}\,l^{2}$. According to Fig. \ref{z=3polymer}, each monomer on the $z$=3 polymer with $g=2$ can be allocated to the following coordinates on the simple cubic lattice ($l=1$) shown in Fig. \ref{g=2polymer}: $\vec{r}_{O1}=(-1,0,0), \vec{r}_{O2}=(0,0,0), \vec{r}_{O3}=(0,1,0), \vec{r}_{O4}=(0,2,0), \vec{r}_{O5}=(0,2,1), \vec{r}_{O6}=(0,2,2)$. From this information, we find quickly $\vec{r}_{OG}=(\frac{-1}{6}, \frac{7}{6}, \frac{3}{6})$ and hence, $\left\langle s_{6}^{2}\right\rangle=\frac{55}{36}$ which is exactly the result that Eq. (\ref{A6}) predicts.\textcolor{blue}{/\hspace{-0.7mm}/}
\begin{figure}[H]
\begin{center}
\includegraphics[width=6.5cm]{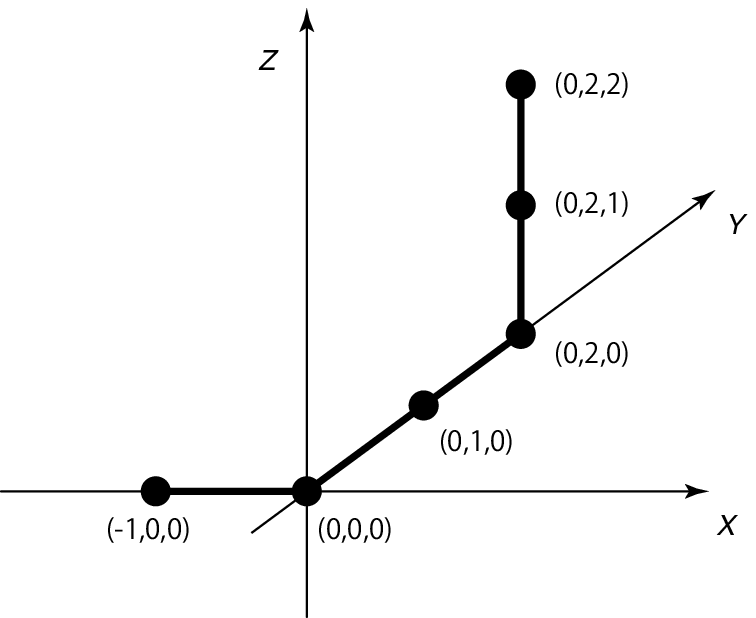} 
\caption{The $z$=3 polymer ($g=2$) having the stretched backbone and the stretched side chains.}\label{g=2polymer}
\end{center}
\end{figure}
\vspace{-5mm}
\end{shaded}

For the freely jointed model, Eqs. (\ref{A2})-(\ref{A5}) give
\begin{equation}
\left\langle s_{g}^{2}\right\rangle_{0, {z=3}}=\frac{1}{6}\frac{(g-1)\left(7g^{5}-4g^{4}-13g^{3}+13g^{2}-2g+1\right)}{g\left(g^{2}-g+1\right)^{2}}\,l^{2} \label{A9}
\end{equation}
As $g\rightarrow\infty$, this reduces to
\begin{equation}
\left\langle s_{g}^{2}\right\rangle_{0, z=3}\doteq\frac{7}{6}\,g\,l^{2}\label{A10}
\end{equation}
which is again proportional to $g$, in accord with the empirical rule (\ref{EF-21}) in the text\cite{Kazumi2}.


\end{document}